\documentclass[useAMS,usenatbib]{mn2e}

\usepackage{psfig}

\title[A batch of $z\sim1$ clusters]{Batch discovery of nine $z\sim1$ clusters using x-ray
and $K$ or $R,z'$ images.
}

\author[S. Andreon et al.]
{S. Andreon,$^1$\thanks{email: andreon@brera.mi.astro.it} 
I. Valtchanov,$^2$ L. R. Jones,$^3$ B. Altieri,$^4$ 
\newauthor
M. Bremer,$^5$ J. Willis,$^6$ M. Pierre,$^7$ H. Quintana,$^8$
\\
$^1$INAF-Osservatorio Astronomico di Brera, Milano, Italy \\
$^2$Astrophysics group, Blackett Laboratory, Imperial College, London, UK\\
$^3$School of Physics and Astronomy, University of Birmingham, Birmingham, UK\\
$^4$XMM-Newton Science Operations Centre, European Space Agency, Villafranca del Castillo, Spain \\
$^5$Department of Physics, Bristol University, Bristol \\
$^6$Department of Physicsand Astronomy, University of Victoria, Victoria, 
	Canada \\
$^7$CEA/DSM/DAPNIA, Service d'Astrophysique, Gif-sur-Yvette, France\\
$^8$Departamento de Astronom\'\i a y Astrof\'\i sica, Pontificia Universidad
Cat\'olica de Chile, Santiago, Chile\\
}
\date{Accepted ... Received ... Version 1}

\pagerange{\pageref{firstpage}--\pageref{lastpage}}
\pubyear{2004}

\begin{document}

\label{firstpage}

\maketitle

\begin{abstract}

We present results of an initial search for clusters of galaxies at
$z\sim1$ and above, using data from 2.9 square degrees of
XMM-Newton images. By selecting weak potentially extended 
X-ray sources with faint or no
identifications in deep, ground-based optical imaging, we have
constructed a starting sample of 19 high redshift cluster candidates. 
Near-IR and $R,z'$ imaging of these fields identified nine of them
as high redshift systems. Six of these were confirmed 
spectroscopically, three at $z\sim1.0$ and the other three in $0.8<z<0.92$ 
range. The remaining three systems have solid photometric evidence to
be at $z_{phot}\sim 0.8, 1.0$ and $1.3$. The present sample
significantly increases the number of such clusters.
The measured
density of $z\ga1$ clusters, after discarding
``low" redshift systems at $z \la 0.92$ is about 1.7 deg$^{-2}$ (with 68 \%
confidence interval equal to $[1.0,2.9]$) for
$f_X \ga 2.5 \ 10^{-15}$ ergs cm$^{-2}$ 
s$^{-1}$ ($[0.5-2]$ keV) and this is a lower limit,
having screened not all potential $z\sim1$ candidate clusters.
Coordinates, x-ray measures and evidence for nine x-ray selected
high redshift clusters is given.
\end{abstract}


\section{Introduction}

Although previous generations of X-ray satellites (notably ROSAT and
Einstein) had the sensitivity to carry out large surveys for
intermediate redshift clusters, they discovered relatively few
clusters at $z>0.8$. The number of the X-ray selected clusters at
$z>0.8$ from these two observatories are twelve, five of which are
at $1<z<1.3$ (e.g. Gioia et al. 1990, 2003; Henry et al. 1997; 
Viklinin et al. 1998;  Nichol et al. 1999;  Rosati et al. 1999, 2004;
Cagnoni et al. 2001; Stanford et al. 2002).

Identifying more X-ray selected clusters at high redshift is a key
step in understanding their evolution and their galaxy
content.  The X-ray luminosity function of clusters shows little or
no evidence for evolution out to $z=0.8$ (Rosati et al. 1998,
Lewis et al., 2002), except for the most luminous (massive) systems
({\it e.g.}  Mullis et al, 2004). Therefore, strong cluster evolution
for more typical clusters is expected to occur at higher
redshifts.

Although studies of the evolution of 
X-ray properties of clusters have been carried out to $z=1.3$
({\it e.g.} Ettori et al, 2004), the conclusions are limited by 
the handful of high redshift systems. Similarly, studies of the
individual galaxies that are found in the cores of intermediate
redshift clusters indicate that there is little evolution in their
stellar populations out to $z=0.8$ (Stanford et al. 1997, Kodama et al. 1997, 
Andreon et al. 2004a).

\begin{table*}
\caption{X-ray properties of clusters}
\label{tab:list}
\begin{tabular}{llrrrlll}
\hline
Long Name & Short Name & $t_{exp}$ & clus+bkg & bkg & $-\log$(P(noise)) & $-\log$(P(ext.)) & Notes\\ 
     & 	& ks & counts & counts & & \\
     (1) & (2) &  (3) & (4) & (5) & (6) & (7) & (8)\\
\hline 
XLSSJ022738.5-031806& XLSSC 003  & 3.96 & 179 & 63  & $>$14  & $>$14 & \\
XLSSJ022708.7-041759& XLSSC 005a & 14.01 & 126 & 54  & $>$14  & $>$14  &  \\
XLSSJ022403.9-041328& XLSSC 029  & 8.37 & 325 & 124 & $>$14 & $>$14  & \\
XLSSJ022157.4-034001& c 	 & 4.32 & 59  & 38  & 3.2  & 2.3 & low surface brightness \\
XLSSJ022253.5-032824& f 	 & 4.96 & 69  & 43  & 4.0  & 1.3 &  \\
XLSSJ022400.5-032526& XLSSC 032  & 2.29 & 61  & 39  & 3.4 & 1.1 & present in two pointings\\
XLSSJ022303.0-043622& h 	 & 9.75 & 68  & 33  & 7.5  & 0.7 &\\
XLSSJ022534.2-042535& l 	 & 12.80 & 63  & 43  & 2.8  & 0.5 & see sect. 3.2\\
XLSSJ022712.0-041835& XLSSC 005b & 14.01 & -- & --  & & --  & blended with XLSSC 005a\\
\hline
\end{tabular}
Counts are for PN only in the [0.5-2] keV band and within the radius
at which the grow curve converged. Hence, count rates derived
from these counts are asymptotic.  
In col. 6, P(noise) is the probability of observing, in the studied area, 
a number of photons larger than quoted in col. 4, given the background (quoted in
col. 5) assumed to be well known (Gehrels 1986, Andreon 2005). Of course,
larger values can be derived considering a smaller area of higher S/N.
In col. 7, P(ext.) is the probability to be a point source, as measured by 
a KS test. 
\hfill \quad \break
\end{table*}

The search for distant X-ray selected clusters is also driven by the
need to obtain more reliable values for the key cosmological
parameters. Despite the recent emergence of the so-called
``concordance'' cosmology, with most of the main cosmological
parameters known with good accuracy, there is still significant
uncertainty in the values of some of them. So, a well-designed and
characterized survey for high redshift clusters can put
complementary and independent constraints and be a key tool for
reliable measurements of some of the cosmological parameters (see
e.g. Refregier, Valtchanov \& Pierre 2002, Borgani et al. 2001, 
Haiman et al. 2001). Individual cluster galaxies can also be
used as a standard
candle in a new cosmological test using the lookback time (Capoziello et al. 
2004).

Here we describe the
results of an initial search for $z\sim 1$ clusters using X-ray
observations of about three square degrees, in order
to prove the feasibility of detecting and studying X-ray selected
clusters at $z>1$ with XMM-Newton plus deep optical or near--infrared
photometry. We deliver position, redshift and basic x-ray quantities
for our nine systems with solid
photometric or spectroscopic evidence to be at high redshift.

Sect. 2 presents the data (X-ray, NIR and $R,z'$ imaging)
and their reduction. Sect. 3 addresses the cluster confirmation using
photometric data, whereas Sect. 4 reports about the spectroscopic
confirmation and focuses on the estimate of the
photometric redshift of the three clusters without spectroscopic
redshift.  Finally, Sect. 5 summarizes the results.

In this paper we adopt $\Omega_M=0.3$, $\Omega_\Lambda=0.7$ and
$H_0=70$ km s$^{-1}$ Mpc$^{-1}$.

\section{The data and the sample}

\begin{figure*}
\centerline{%
}
\centerline{%
}
\caption[h]{X-ray growth curves for sources, in order of likelihood of
extension (i.e., as in Table 1).  
Source XLSSC 005b is missing because contaminated by XLSSC 005a.  The solid
curve is the PSF plus background profile.}
\label{fig:curves}
\end{figure*}

\begin{figure*}
\centerline{%
}
\centerline{%
}
\caption[h]{X-ray [0.5-5] photon maps with superposed X-ray contours
for high redshift clusters XLSSC 005a,b, XLSSC 029, 
{\it c, f, XLSSC 032, h, l} (from left to right, and top to down).   
The grey-scale images are 3x3 arcmin in size with 8 arcsec pixels.
Contours of adaptively smoothed emission are logarithmically spaced,
starting at 2$\sigma$ significance above the background in the
smoothed image with (logarithmic) step of about 0.1. All features visible in the contours are significant
at the 99\% confidence level. N is top, E is to the left.}
\label{fig:images}
\end{figure*}

\begin{figure*}
\centerline{%
\vbox{
}}
\caption[h]{
Three colour ($Rz'K_s$) images of high redshift clusters 
with superposed X-ray contours. N is top, E is to the left.
For the special source {\it l} we also marks some spectroscopical
confirmed member with small white circles.
The ruler is 1 arcmin wide. No three colour image of XLSSC 003 
is available, because unobserved in $R,z'$.}
\label{fig:colour}
\end{figure*}



%

\subsection{X-ray data and sample selection}

The sample of candidate clusters is drawn from the first
connected 2.9 degrees of the 
XMM Large-Scale Structure Survey 
(Pierre et al. 2004, Andreon \& Pierre 2003) 
available at the time of NIR observations (November 2002). 
This comprise 26 individual contiguous XMM-Newton pointings
with exposure times mostly of 10 ks and some deeper ones of  
typically 20 ks (see
Pierre et al. 2004). 

In this pilot search, our aim is to demonstrate that we can found
$z\sim1$ clusters efficiently 
without being concerned with the
precise knowledge of the cluster selection function, which indeed is
needed for cosmological applications and is the focus of Bremer et al. (in
preparation).
A massive Coma-like cluster ($kT \approx 7$ keV) at $z=1$ will
result in XMM-Newton detection of $\sim 600$ photons in 10 ks on the
PN detector, while medium to low-mass cluster ($kT \approx 3$ keV,
like A1060 for instance) will give $\sim 70$ counts. Therefore, 
both should be easely detected in our exposures.

The data were analysed and a list of possibly extended sources were
generated in the manner outlined in Valtchanov, Pierre, \& Gastaud
(2001). Their classification procedure
efficiently excludes point-like sources although it leaves some
complex cases (mostly objects of low signal-to-noise or
contaminated by a nearby source), as shown by the grow curve
analysis described below.

We have excluded all X-ray 
sources at more than 12 arcmin off-axis, in order to avoid low
sensitivity and regions with distorted PSF.

In the second step, we use optical $I$-band imaging of the complete
field from the CFH12K camera on CFHT (the VIRMOS-VLT Deep survey,
MacCracken et al. 2003).  The images were of
sufficient depth ($I\ga23$ mag) to easily recognize  a $z\la0.8$
cluster, but insufficient to detect, at high significance, higher
redshift clusters, because of their low contrast with respect to the
background.  For each X-ray source, the $I$-band image was examined
for the presence of any obvious galaxy overdensity. If found,
the X-ray source is discarded.
A few more sources are discarded when they match obvious
low redshift targets, for example a resolved (nearby) galaxy in the
optical. From the resulting shortlist we have chosen 19
examples, putting priority on X-ray sources spatially coincident
with optically faint galaxies, suggestive of a 
$z\sim1$ cluster.
Nine of these examples have been confirmed at present as high
redshift systems, and here we focus on them.
Table~\ref{tab:list} lists a few basic X-ray features. The faintest cluster
has about 20 PN net counts in a 13 ks exposure, 
corresponding to a flux of about $2.5 \ 10^{-15}$ ergs cm$^{-2}$ 
s$^{-1}$.

The probability of extension listed in Table 1 is measured
{\it after} the cluster candidate choice in
the following way: for each source we construct the X-ray growth
curve (also called integral profile) using only the PN detector
photons in the $[0.5-2]$ keV energy band. The candidate cluster
location, extent, background region and possible overlapping sources
that need to be excluded from the analysis are defined from the
adaptively smoothed image of the three XMM-Newton detectors
(MOS1+MOS2+pn).  We use the XMM-SAS task {\tt asmooth} with
signal-to-noise ratio of 5. Then the list of photons from the source
region is extracted with {\tt xmmselect} and the standard selection
of good PN events. The background is measured in an adjacent
annulus. The extracted X-ray growth curve is then compared to that
expected for a point source at the the same angular distance from
the centre using the Kolmogorov--Smirnov test.

Figure~\ref{fig:curves} shows the X-ray growth curves for 8 out
of 9 high redshift systems. The remaining source, XLSSC 005b, 
falls too near to  a bright source for a reliable extension measure.
The observed profile is compared to the XMM-Newton point spread function
(PSF) encircled energy fraction (the King $\beta$-model, Ghizzardi 2001)
convolved with the measured background.  Note that the PSF EEF varies
insignificantly with the off-axis angle at all studied off-axis angles
(Ghizzardi 2003). The KS test applied to these pairs of curves gives the
likelihood of extension listed in Table 1. Three sources are clearly
extended, whereas two faint sources show no evidence for 
extension. The latter sources are in our list because of the conservative
(and appropriate) choice of the Valtchanov et al. (2001) of keeping ambiguous 
cases in the sample.

Figure~\ref{fig:images} shows the combined (MOS1 + MOS2 + pn) XMM
images in $[0.5-5]$ keV energy band,  
for the X-ray sources that turned out to be photometrically
or spectroscopically confirmed high redshift clusters. 
The images are corrected for vignetting and chip gaps using the
appropriate exposure maps and over-plotted with the contours from
the adaptive smoothing. In some cases (eg. {\it h}) point-like
X-ray sources overlap the extended X-ray emission and might
confuse the measurement of the extent probability. In all these
cases, the point source is offset from the centroid of  the
target emission, so that its presence will increase the
extent probability.

\subsection{NIR observations}

In order to confirm the candidates as clusters we need 
images in a redder band than the existing I band CFH12k observations. 
Galaxies at the
centres of distant clusters generally have colours of passively
evolving ellipticals and are consequently red objects. At $z\sim 1$,
galaxies close to the centres of clusters typically have $R-K >4$
(Vega), a magnitude or more redder than the general faint galaxy
population. Thus, high redshift clusters, otherwise difficult to
identify in optical imaging, are detectable as clear overdensities of
faint ($K>17$) galaxies in near IR imaging.

$K_s$-band images of the sample were acquired with SOFI (Moorwood,
Cuby \& Lidman, 1998) on the 3.5m NTT at La Silla on November 19/20
and 20/21, 2002.  SOFI is a near-infrared imager equipped with a
$1024\times1024$ pixels Rockwell "Hawaii" array and was used in its
large field mode, with a field of view of $5\times5$ arcmin and a
pixel size of $0\fs289$.

All our 19 targets were exposed for 30 min, except a target 
not presented in this paper, 
resulting from the coaddition of 
many short jittered exposures.
Photometric calibrations were obtained by observing a series of NICMOS
infrared standard stars (Persson et al., 1998).  The adopted value for
the atmospheric absorption is 0.05 magnitudes in $K_s$.  The derived
zero-point from all standard stars was found to be constant throughout
the two nights with a scatter of 0.01--0.02 mag.

All images were flat-fielded by a special flaton-flatoff frames fully
described in the SOFI manual (see also Andreon 2001).  The {\tt
  IRAF/DIMSUM} external package developed by Eisenhard et al.  (see
info at the IRAF www site) was used to produce accurate sky subtracted
images from the dithered observations.
Object detection has been performed by using SExtractor (Bertin \&
Arnouts 1996).

On the composite image, the seeing varied from 0.61 arcsec FWHM
to 0.85 arcsec.
In order to build a galaxy sample, all obvious stars have been removed
from the sample by using the SExtractor stellarity index. The
catalogue turned out to be complete down to magnitude $K_{s}\sim
20$ measured in 2 arcsec apertures.

Figure~\ref{fig:colour} shows the three-colour ($Rz'K_s$) images of
eight out of nine high redshift clusters, superposed by the X-ray contours
shown in Figure~\ref{fig:images}. Three-colour image of 
the remaining system is missing
because unobserved in $R$ and $z'$. Note the diversity of 
the optical-NIR appearance of the clusters:
{\it h} is quite compact, whereas source {\it c} extends out to
the nearby extended source in the W; XLLSC 005b appears, at first
sight, to have a galaxy population more variegate in colour than
XLSSC 029.

\subsection{$R$ and $z'$ observations}

$R$ and $z'$ images have been taken at the 4m CTIO telescope during
two observing runs, in August 2000 and November 2001, with the Mosaic
II camera.  The Mosaic II is a camera of 8k$\times$8k pixels
with 36 $\times$ 36 arcmin field of view.  The observing time was
typically 1200s in $R$ and 2$\times$750s in $z'$. The seeing in the
final images was between 1.0 and 1.4 arcsec in the November 2001 run,
and 0.9 to 1.0 arcsec in the August 2000 run. The useful nights of the
two observing runs were photometric, however images taken during the
2001 run have been acquired at bright time (about 70 \% illuminated
moon), making them shallower than images taken in the previous
(dark time) run. Despite being observed during bright time, all three
spectroscopically confirmed clusters at $z\sim1$ can be 
identified (and detected) in these images, as Figure 3 shows. 
Images are reduced as usual and more details can be found in Andreon
et al. (2004a). {Typical completeness magnitudes are $R=25$, $z'=23.5$ mag
($4\sigma$) in a 3 arcsec aperture}.

We used SExtractor (Bertin \& Arnouts 1996) for object
detection and characterisation. The colours were measured inside 3
arcsec apertures. We have adopted a larger aperture in $R$ and $z'$ than in
$K_s$ to account for the  worse seeing of the optical data.
Obvious stars (those with SExtractor stellarity index larger than
0.95) were discarded from the galaxy sample.

$R$ and  $z'$ images are available for 13 out of the 19 selected targets
and, in particular, for 8 out 9 our photometrically/spectroscopically 
confirmed high redshift clusters. Figure 3 is produced using these
data.

\begin{figure}
\psfig{figure=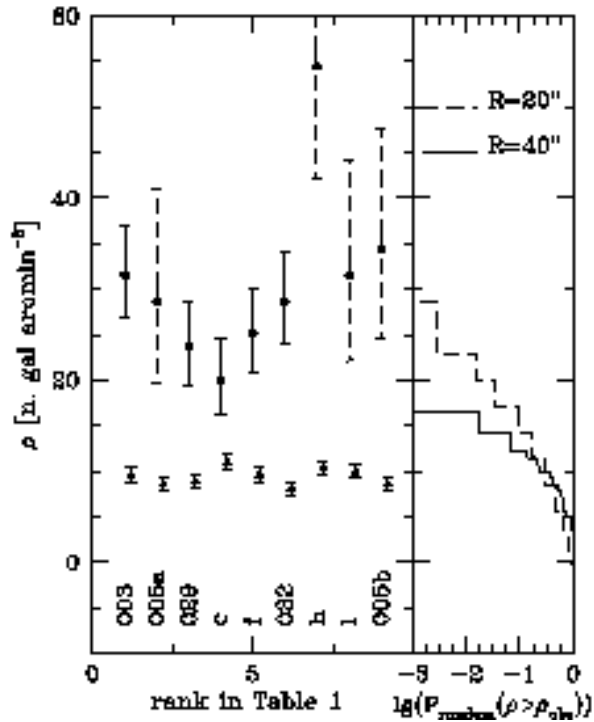,width=8truecm}
\caption[h]{{\it Left panel:} Surface density of galaxies {\it vs}
  rank in likelihood of extension. For each X-ray source
  there are two points: the dot indicates the galaxy surface density
  inside the X--ray isophote, while the triangle the surface density
  in the remaining of the SOFI frame (typically 20--25
  arcmin$^2$). Detection of a possible cluster requires a
  significantly higher surface density within the X-ray isophote than
  in the remaining of the frame. Error bars are 68 \% confidence
  intervals. For the source {\it l}, the optical position is taken. 
  {\it Right panel:} Probability of randomly finding a
  given, or larger, surface density taking the data set as a whole,
  i.e.  probability that the overdensity is a statistical fluctuation.
  There are two curves with two line styles, each one referring to
  points with error bars drawn with the same style.} 
\label{fig:dens}
\end{figure}

\section{Results}

\subsection{Cluster confirmation using the $K_s$-band data}

In order to identify an extended X-ray source as cluster of
  galaxies we look for the presence of a galaxy overdensity in the region
  where the X-ray source is detected. This overdensity must be
  significant enough so to be ruled out as a chance fluctuation.
For all 19 cluster candidates X-ray sources in the SOFI field of view we have
counted the surface density of galaxies brighter than $K_{s}=20.5$
inside the area where an excess of X-ray photons is detected
(i.e. within the X-ray isophotal area), approximated as a circle of
equivalent area.  X-ray sources have an equivalent radius
of 40 or 20 arcsec. $K_{s}=20.5$
is slightly deeper than the completeness limit, and has been
adopted in order to make full use of the data.

The left panel of Figure~\ref{fig:dens} shows the galaxy density,
$\rho$, for the nine high redshift clusters, with the
Poissonian 68 \% confidence intervals (Gehrels 1986, Andreon 2005).
Sources with solid/dashed error bars are 40/20 arcsec in radius.

The triangles in the left panel of Figure~\ref{fig:dens} show the
average density (and its Poissonian 68 \% confidence interval) 
measured in each SOFI
pointing, after excluding the image borders (where the exposure time
is lower) and a circle twice larger than the X-ray source.

In order to accurately measure the statistical significance of the
galaxy overdensity, taking into account any possible
contribution related to galaxy structures other than our 19
putative clusters, we proceed as follows: we randomly
put square apertures of the same area as the actually detected
cluster on our SOFI images,
and we count how many times we detect a density, inside the
apertures, higher than $\rho$. 
As before, we exclude image borders and a circle
twice as large as the X-ray source radius. The right panel of
Figure~\ref{fig:dens} shows the frequency of observing by chance a
number density $\rho$ or higher. A large negative number means
that the overdensity is unlikely to be a background fluctuation.  The
curve is not smooth because the galaxy number density can take
only discrete values.  This approach does take into account the
angular correlation function of galaxies and is preferable than
assuming a Poissonian distribution of the galaxy counts.

Following our analysis we have found:

\begin{itemize}

\item Four potentially extended X-ray sources with scale of 20 arcsec
  (XLSSC 005a, {\it h}, {\it l} and XLSSC 005b) are spatially coincident
  with galaxy overdensity in $K_s$ which was seldom observed in the
  control field (649 independent
  lines of sight, the 18 SOFI images minus twice the area occupied by
  the targets). Their statistical significance is therefore larger
  than 0.999. Note, however, that the center of source {\it l}
  has been moved away from the x-ray center by about 40" (see sec 3.2).

\item Five extended sources with scale of 40 arcsec (XLSSC 003, XLSSC
  029, {\it c},  {\it f}, XLSSC 032), are spatially
  coincident with galaxy overdensity in $K_s$ which was never
  observed in our control field (51 independent lines of sight).  The
  statistically significance is larger than 0.98 but it is worth to
  note that we cannot quote a larger confidence level because of the
  limited area of the control field, not because of the ``low signal"
  of the detection.

\end{itemize}

All listed statistical significances include contribution from
clusters/groups unrelated to the X-ray emission (line of sight
projections), i.e. the statistical analysis takes into account the
possibility that the X-ray emission is, {\it by chance} on the same
line of sight of the cluster but at different redshift (i.e. the X-ray
photons does not come from the NIR identified cluster). The
detected galaxy overdensity is not due to the cosmic variance of
galaxy counts, because it is not shared by the whole SOFI field of
view where the clusters are detected, 
as the left panel of Figure~\ref{fig:dens} shows
(compare the triangles and the circles).  The stated statistical
significance holds for the assumed aperture.  An optimal aperture 
will only increases the statistical significance of the NIR cluster
detection.

Concluding, we have found galaxy overdensities with statistical
significance larger than 0.98 in direction of all presented nine
cluster candidates, with the caveat of the spatial offset of 
source {\it l}. 

\begin{figure*} 
\psfig{figure=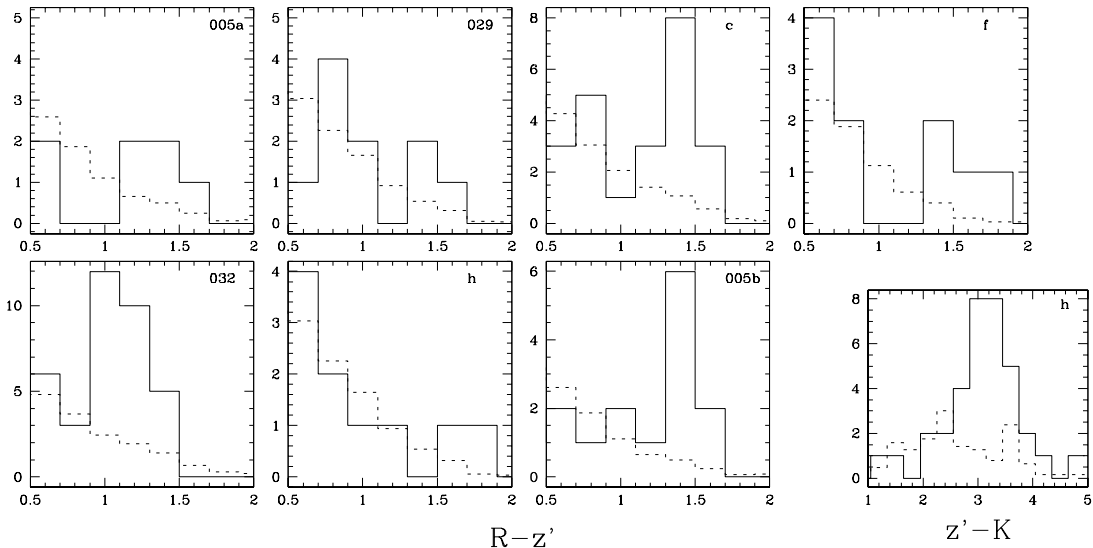,width=15truecm} 
\caption[h]{ $R-z'$ colour distribution of the galaxies brighter than
  the $z'$ completeness limit within a circle of 3.14 arcmin$^2$ area
  centered on the X--ray cluster centre (solid histogram) and in a
  control field given by the remaining field of view of the 36
  $\times$ 36 arcmin pointing in which the cluster is observed, 
  normalised to the same area. The bottom-right panel shows
  a preliminary colour distribution in $z'-K$ of cluster {\it h} based
  on recently received (Gemini $z'$) data.}
\label{fig:rz}
\end{figure*}

\subsection{Cluster confirmation using $R-z'$ colour--selection}

Clusters of galaxies up to $z\sim1.3$ should be detectable in colour
space using $R-z'$ (e.g. Gladders \& Yee 2000). We have developed
a method to detect clusters inspired by, but different from Gladders 
\& Yee (2000) and described in Andreon et al. (2003b,
2004a,b). Shortly, the method takes advantage from the observed fact
that most of galaxies in clusters share similar colours, while
background galaxies have a variety of colours, both because they are
spread over a larger redshift range and because the field population
is more variegate in colour than the cluster one, even at a fixed
redshift. In practice, the method looks at spatially localized galaxy
overdensities of similar colour.

Figure 5 shows the observed colour distribution (solid line) and the
expected contribution unrelated to the cluster (dashed histogram) measured in a
control field chosen as described in the figure legend.  The cluster
contribution is given by the area between the solid and dashed histograms, and
its significance can be easely computed (this is the standard $(on+off) \
-(off)$ problem discussed several times in literature including, say, Loredo
1992). In Figure 5 we considered galaxies within a fixed distance from the
x-ray measured center, and we arbitrary chose the amplitude of the colour bin.
In the detection algorithm, the location and scale of the detection kernel in
the spatial and colour space are optimized. The significances quoted in Table 2
comes from the optimization and are larger than one may derive from reading
Figure 5 and straightforward application of statistics. These confidences are
quoted with more decimals than for confidences determined from $K$ band data
because of the large number of  independent line of sights  sampled with the
huge control field available for $R-z'$ data.

Table 2 shows that in for
6 out of 8 candidates having both NIR and $R,z'$ data, the $K$ band
and $R-z'$ detection agree on 
the detection of a galaxy overdensity spatially
coincident with the X-ray source.  There is an apparent disagreement
in just two cases:

\begin{itemize}
\item 
source {\it h}, which brightest cluster galaxy is faint (see
  Fig. 3, Section 4.1 and Table 2) and the colour
  of the reddest galaxies in the cluster direction is very red
  (Figure~\ref{fig:rz}). During the paper revision, deeper 
  $z'$ imaging has been taken at Gemini and the reason for the non 
  $R-z'$ detection
  clarified: the cluster is now colour detected at 
  $z_{phot}\sim1.3$ (estimated from the peak of the colour distribution) 
  but using $z'-K$ as colour (Figure 5, bottom--right
  panel). The non-detection
  in $R-z'$ comes from the too large distance of the cluster,
  associated to the unfortunate location of the cluster near
  a bright star whose diffraction spike crosses the cluster centre 
  in the old $z'$ data (see Fig. 3).

\item 
Source {\it l} is a complex case. At the exact
location of the presently determined
x-ray center there is no overdensity of galaxies in $K$-band 
or in colour
space, reason why the source is undetected in both $K$ and
$R-z'$ colour. However, there is a galaxy overdensity
in the $K$-band about 40 arcsec west (about 300 kpc at $z\ga1$),
detected as such in the $K$-band (and reported in Fig 4), spectroscopically confirmed
(four central galaxies have $z=0.921$ within 500 km/s, as well
some other galaxies scattered over the field, see Fig 3), 
but undetected in $R-z'$, probably
because the extreme poorness of the group and because
its spatial distribution is more compact than the smallest
scale used in the cluster detection (30 arcsec). While the existence
of an x-ray source and of a cluster of galaxies is beyond doubt,
the identification of the $K$-band and
spectroscopically confirmed cluster with the x-ray source
is only possible, because the small, but not negligible, 
distance (300 kpc) between the
optical and x-ray sources. Another possibility is
that a weak emission centered on the optical
cluster center, just below the detection threshold
may be there, but we have detected only the portion of it
contaminated by a nearby X-ray source.

\end{itemize}

Therefore, to confirm the high redshift clusters in our sample,
the detection in $R-z'$ is almost as efficient as the NIR one. $R-z'$
performance is not expected to continue to be as good at $z \gg 1$,
and redder/deeper data will be needed, as for source {\it h}.

\begin{table*}
\caption{Results}
\label{tab:results}
\begin{tabular}{llllclll}
\hline
Short name  & NIR det?      &  $R-z'$ detected?		&   $z_{spect}^{(1)}$ & $R-z'$ & K(BCG)  & $z_{phot}$  &Notes \\
\hline
XLSSC 003    &     $>$0.98 c.l.&   Out of f.o.v.	  & 0.84	  &	   & 16.9    &  &   \\
XLSSC 005a    &     0.999 c.l   &   $>$0.999995 c.l.			&	   1.0    &	
  & 18.1    &  & blended with XLSSC 005b \\
XLSSC 029    &	 $>$0.98 c.l.&   $>$0.999995 c.l.	& 1.05	        &  1.5   & 17.3	 
 &  &   \\
{\it c}    &	 $>$0.98 c.l.  &   $>$0.999995 c.l.	  &		  &  1.3   & 16.6    & $0.8-0.9$ & \\
{\it f}    &	 $>$0.98 c.l.&    $>$0.999995 c.l.	  &		  &  1.5   & 16.3    & $\sim1$ &  uncertain BCG ID \\
XLSSC 032    &	   $>$0.98 c.l.&   $>$0.999995 c.l.	  & 0.81          &  1.2   & 16.3    &  &    \\
{\it h}    &	   0.999 c.l. & NO			  &		  &	   & 17.5    &
$\sim1.3$ & $z'-K$ detected    \\
{\it l}    &	   0.999 c.l. &   NO			  & 0.92	  &	   &	     &  & optical-X uncertain
identification    \\
XLSSC 005b    &     0.999 c.l.  &  $>$0.999995 c.l.	  & 1.0		  &  1.45  & 18.1    &  &     \\
\hline
\end{tabular}
\hfill \break
$^{(1)}$ Valtchanov et al. (2004), Willis et al. (2005 and private communication)
and Sprimont et al. (private communication), to be
published in the XMM-LSS cluster catalogue.
\hfill \quad \break
Note: In col. 3, "NO" means undetected at the 97 \% confidence level. 
\hfill \quad \break
\end{table*}

\section{Redshift determination}

\subsection{Spectroscopic redshift}

Table~\ref{tab:results} lists the spectroscopic redshift determined
during fall 2002 (the time of the NIR campaign) or later.
Three clusters turned out
to be at $z\ga1$. Two, instead, are ``low'' redshift (but $z>0.8$!)
and escaped our (eye-based) cluster identification procedure,
similarly to the cluster possibly identified with
the source {\it l} at $z=0.92$\footnote{The redshifts were measured 
and kindly provided by 
the VIRMOS consortium}. Here ``spectroscopically confirmed" means 
that at least 3 galaxies
falling inside the x-ray emitting region have measured spectroscopic
redshifts within 1000 km/s or less, following standard practice
(e.g. Mullis et al. 2003; B{\" o}hringer et al. 2004), with the mentioned
exception of the special source {\it l}.

XLSSC~005a and XLSSC~005b are two parts of a complex structure at
$z\approx1$.  The structure has thickness of about $10^4$ km/s
(Valtchanov et al. 2004), and hence the two objects are unlikely to
be part of one single cluster and we count them as two separate
entities. The superposition of the two structures both
in the plane of the sky and in redshift, together to a limited sampling
in redshift space (eleven measures inside the central arcmin),  
force us to give a coarse redshift for the two sources in Table 2.

\subsection{Redshifts assuming BCGs as a standard candle}
\label{sec:z:bcg}

The Hubble diagram (i.e. redshift vs magnitude) for brightest cluster
galaxies (BCGs) is a classical cosmological tool (e.g.  Sandage
1995). The dispersion of the BCG magnitude in the $K_s$-band at a
fixed redshift is about 0.22 mag in homogeneously selected samples
(Collins \& Mann 1998; Brough, Collins, Burke, Mann, \& Lynam, 2002),
but easily become twice as large when an heterogeneous sample is taken
(e.g. including BGCs listed in Stanford et al. 2002 or
Aragon--Salamanca et al. 1993). For the assumed cosmology, a 0.22 mag
dispersion correspond to $\sigma_z=0.1$ in redshift. 
Mullis et al. (2003) quote a slightly better figure ($\sigma_z=0.1$ 
90 \% confidence, instead of 68 \% confidence) for a large X-ray selected
sample, but at $z_{median}=0.25$.

Table~\ref{tab:results} lists the SExtractor isophotal corrected
K-band BCG luminosity for all our 9 targets, but the special source
{\it l}. Six of the objects have a spectroscopic redshift and
allow us to empirically calibrate the Hubble diagram using our own
data.  Out of the 3 candidate clusters without $z_{spec}$ and spatially
coincident with a galaxy overdensity in NIR, one have $K_s(BGC)$
fainter than, or similar to, our $z_{spec}\sim1$ clusters. Two,
instead, are quite brighter suggesting $z_{phot}\sim0.8$, inferred
from the comparison with our clusters with $z_{spec}$.

Photometric redshifts estimated from the luminosity of the BCG are
prone to be underestimated, because of the possibility that the galaxy
identified as the BCG is instead an interloper: in such a case the
actual BCG would be fainter and the cluster more distant than claimed.

\subsection{Redshift from colours}

Assuming that the colour of the reddest galaxies in clusters evolve
passively with $z$, we can infer the (photometric) redshift of
the cluster from the colour of the red sequence. The method has been
used, and confirmed, for 5 X-ray selected clusters at $0.1<z<0.5$
(Puddu et al. 2001), in 140 cluster at $0.06<z<0.30$ (Andreon 2003),
in 19 clusters at $0.3<z<0.9$ by Stanford et al. (1998), in 18
clusters $0.3\la z \la 1$ by Andreon et al. (2004a).

Instead of relying on galaxy evolutionary models, we can afford a
direct approach: since we known, by our own observations of XLSSC 005b
and 029, the colour of the main population of a cluster at $z=1$
($R-z'\sim 1.45-1.5$ mag, see Table~\ref{tab:results}), we can derive
the other photometric redshifts by comparison with the reference one.

There are only two clusters colour detected lacking
a spectroscopic redshift:
one (source {\it f}) has
$z_{phot}\sim1$ and one (source {\it c}) has $z_{phot}\approx0.8-0.9$. 
There is a good agreement between  the photometric redshift
($z_{phot}$) derived by colour and by the BCG luminosity for source {\it c},
but disagreement for source {\it f}, whose $R-z'$ colour
suggest $z\ga1$, while its BCG suggest $z\la0.8$. For such a cluster
it is possible that the galaxy identified as BCG is, instead, an
interloper.

Source {\it h}, instead, is NIR detected but not colour detected
in the old data. The
non-detection of the source {\it h} in $R-z'$ and the red colour of the
reddest of its galaxies shown in Figure~\ref{fig:rz}, both imply
$z>1$, in good agreement with the value estimated from the colour
of the red sequence from new $z'-K$
data ($z_{phot}\sim1.3$).

\section{Summary and discussion}

We have selected 19 X-ray potentially extended sources,
without
obvious counterpart in deep optical images, and hence are 
candidates for $z\ga 1$ clusters.   
All 19 objects were observed with SOFI at NTT, while 14 of them were
in the field surveyed in $R,z'$ at CTIO.

NIR/$R,z'$ observations photometrically confirm night of them as
genuine high redshift clusters: they are detected as NIR galaxy overdensity,
their galaxies share similar enough colours that the cluster is
detected by the
$R-z'$ ($z'-K$, in one case) colour detection algorithm, their brightest galaxy members have
luminosity compatible with $z\sim1$ and the galaxies on the
colour magnitude relation (the colour of detection) have the right
colour to be early-type galaxies at $z\sim1$.  Spectroscopical
observations already confirmed three of them as spectroscopic $z\sim1$
clusters, and three more as $0.81<z_{spec}<0.92$ clusters. The
three remaining clusters have $z_{phot}=0.8-0.9, 1.0$ and $1.3$.

Our success in identifying high redshift clusters 
does not come unexpected. However, we show that expectations are
satisfied with data. For all identified systems
we give position, basic x-ray quantities (fluxes for
example, can be computed from numbers quoted in Table 1), 
solid evidence about their photometric (high) redshift and,
in six cases out nine, spectroscopic redshift. With the current
work the number of high redshift clusters with x-ray emission
has approximatively doubled.

Cluster ({\it h} has very low likelihood 
of extension, still it is a cluster, suggesting caution in
using likelihood of extension at low x-ray counts.

About half of our 19 X-ray sources are not identified as clusters
of galaxies (and not reported here individually). 
These sources should not be considered as
a failure: they are unidentified sources at present.  However, the
identification may change with new data, which probe, for example, a
different redshift range ($z\ga1$ with UKIDSS/Spitzer data) or which increases
the significance of the NIR/$R-z'$ detection, or which allow a full
three dimensional sampling.

Overall, there are 5 clusters with solid evidence to be at
$z\ga1$ (we neglect clusters at $z=0.81,0.84,0.92$ and the
cluster at $z_{phot}=0.8-0.9$)  
in the inspected 2.9 deg$^2$. Three of them are spectroscopically
confirmed to be at $\ga1$.
The number density of observed
clusters at $z\ga1$, implied from this study, is about 1.7
deg$^{-2}$ for clusters with $f_x \ga 2.5 \ 10^{-15}$ ergs cm$^{-2}$ s$^{-1}$
$[0.5-2]$ keV (the flux of the faintest considered source).
The 68 \% confidence interval (Gehrels 1986, Andreon 2005), 
assuming a Poissonian
probability distribution function, is $[1.0,2.9]$. This estimate is a
lower limit, because not all sources in the considered area have been
scheduled for NIR observations. {We leave the discussion of the 
volume density of clusters at $z>1$ and selection function to Bremer
et al. (2005)}.

Colour detection and NIR overdensity achieve similar performance in
confirming potentially extended X-ray sources deemed to be at $z \approx
1$. Therefore, we are currently considering a cross-correlation between
colour detected and X-ray detected clusters, without the restriction to
sources with NIR images (i.e. the targets of this study) to enlarge the
$z\sim1$ cluster sample. The cross-correlation with X-ray is compelling 
if clusters
with deep potential wells are to be selected and discriminated against
structures of larger size (and similar mass) and from superposition
of groups nearby in space, but not part of a single halo.

\section*{Acknowledgements}

We acknowledge the VIRMOS consortium for the redshift of cluster {\it l}.
The authors thank all the XMM-LSS consortium members 
for stimulating conversations, and Sergio dos Santos
for its early help at the start of this project. Referee 
suggestions helped to improve the paper presentations, and we thank
him. H. Quintana is thankful for partial support to the FONDAP
Astrophysics Center. This paper is based on
observations obtained with XMM, with ESO (prop. 70.A-0733), 
Cerro-Tololo (prog. 0295 and  0316) and, in part, 
Gemini (GN-2004B-Q-55) telescopes.

\bsp

\label{lastpage}

\end{document}